# Exact Results for the Residual Entropy of Ice Hexagonal Monolayer

De-Zhang Li[1], Wei-Jie Huang[1], Yao Yao[1] and Xiao-Bao Yang[1*]

[1] Department of Physics, South China University of Technology, Guangzhou 510640, China.

[*] Corresponding authors. Correspondence and requests for materials should be addressed to X.-B. Y. (email: scxbyang@scut.edu.cn).

ORCID of the authors:

De-Zhang Li: 0000-0001-8039-449X

Wei-Jie Huang: 0000-0003-1881-4099

Yao Yao: 0000-0002-4073-8240

Xiao-Bao Yang: 0000-0001-8851-1988

## Acknowledgements

This work was supported by Guangdong Basic and Applied Basic Research Foundation (Grant No. 2021A1515010328), Key-Area Research and Development Program of Guangdong Province (Grant No. 2020B010183001), and National Natural Science Foundation of China (Grant No. 12074126).

# Abstract

Since the problem of the residual entropy of square ice was exactly solved, exact solutions for two-dimensional realistic ice models have been of interest. In this paper, we study the exact residual entropy of ice hexagonal monolayer in two cases. In the case that the external electric field along the z-axis exists, we map the hydrogen configurations into the spin configurations of the Ising model on the Kagomé lattice. By taking the low temperature limit of the Ising model, we derive the exact residual entropy, which agrees with the result determined previously from the dimer model on the honeycomb lattice. In another case that the ice hexagonal monolayer is under the periodic boundary conditions in the cubic ice lattice, we employ the six-vertex model on the square lattice to represent the hydrogen configurations obeying the ice rules. The exact residual entropy in this case is obtained from the solution of the equivalent six-vertex model. Our work provides more examples of the exactly soluble two-dimensional models.

## 1. Introduction

Research of ice system has been an important theme in the fields of physics and chemistry for a long time. Since in 1930s the ice rules [1, 2] were proposed to explain the non-zero entropy of ice at low temperatures [3, 4], the solution of this residual entropy in various ice systems has been an important and interesting problem in statistical physics and mathematics. The residual entropy arises from the hydrogen configurations obeying the ice rules as $S/k_B = \frac{1}{N_{H_2O}} \ln W = \ln w$, where $W$ is the number of hydrogen configurations, $N_{H_2O}$ is the number of H$_2$O molecules and $w = W^{1/N_{H_2O}}$. We list some famous early studies of this problem: Pauling's mean field approximation $w = \frac{3}{2}$ for four-coordinated ice systems [2], DiMarzio and Stillinger's matrix method for square ice and three-dimensional ice [5], Nagle's series expansion for square ice, ice Ih (hexagonal ice) and ice Ic (cubic ice) [6-8], Lieb's exact result from transfer matrix method for square ice $w = \left(\frac{4}{3}\right)^{\frac{3}{2}}$ [9, 10]. With the developments of the high-performance computational technology, there have been various numerical simulations [11-22] and theoretical evaluations based on computer [23-26] for this problem.

Among the works of the residual entropy problem of ice systems, our interest is the exact solution of the two-dimensional ice models in this article. Since the simple one-dimensional Ising model was solved in 1920s [27], the studies of exact solutions for various statistical models have been of interest [28, 29]. Most of the exactly solved statistical models are in one dimension and two dimensions. For the two-dimensional Ising model on the square lattice, Onsager derived the solution for the case that without an external field [30], and Lee and Yang gave the solution for the case that in an imaginary field [31]. The solutions of the Ising models on the honeycomb lattice [32, 33], the triangular lattice [34] and the Kagomé lattice [35, 36] were also obtained. In particular, the residual entropies from the frustration of the triangular model [34] and the Kagomé model [36] were exactly solved. It is clear that, the residual entropy of ice systems and that of the Ising models on the frustrated lattice have a close relation [37-39]. In 1967, Lieb published the famous solution of square ice [9, 10], which motivated a lot of studies of vertex models like the six-vertex [40-46], eight-vertex [47-52] and sixteen-vertex models [53-60]. The square ice can be seen as a special case of the six-vertex model, and the exact residual entropy can be rederived from the low temperature limit of the Ising model with crossing and four-spin interactions on the checkerboard lattice [61-63]. These two-dimensional statistical models as well as a few others like the dimer model [64-73], the monomer-dimer model [74-79] and the hard hexagon model [80], either solved or unsolved, are briefly reviewed in the Introduction of [60].

In this article, we focus on a two-dimensional ice model, namely, the ice hexagonal monolayer. Two-dimensional ice models have attracted much attention [24, 25, 81, 82] since the research of square ice. Unlike the square ice, ice hexagonal monolayer is a realistic structure in three-dimensional ice such as ice Ih and ice Ic. Although the exactly solved three-dimensional models are rare [83-85], research of realistic two-dimensional structure may provide new insights into the physics of real ice. The exact result of the residual entropy of ice hexagonal monolayer is obtained in two cases. In Sec. 2, we consider the case that in the presence of an external electric field along the z-axis. The exact residual entropy is derived from the low temperature limit of an equivalent Ising model on the Kagomé lattice, and shown in agreement with the solution of dimer covering on the honeycomb lattice. In Sec. 3, the exact residual entropy is solved in the case that the ice hexagonal monolayer is under the periodic boundary conditions in ice Ic. In this case we employ the mapping of the hydrogen configurations into an exactly solved six-vertex model. Discussions and conclusions are summarized in Sec. 4.

## 2. Ice Hexagonal Monolayer in an External Field along the Z-axis

Ice hexagonal monolayer consists of the armchair $(H_2O)_6$ rings. All the oxygens are three-coordinated, hence there are three hydrogen bonds and one dangling bond along the z-axis for each oxygen in the layer. $\frac{3}{2}N_{H_2O}$ hydrogens are in the hydrogen bond network and $\frac{1}{2}N_{H_2O}$ hydrogens are in the dangling bonds. In each hydrogen bond there are two possible positions for a hydrogen, and in each dangling bond there are also two possibilities: one hydrogen or none. It is natural to introduce the mapping of these two possibilities of each bond into the value +1/-1 of an Ising spin. To do this, we should first define the standard direction (+1) of the bonds, which is shown in Fig. 1. Then the bond configurations can be mapped into the spin configurations of the equivalent Ising model, with the value +1/-1 of each spin representing the direction of the corresponding bond. We show the equivalent Ising model in Fig. 2.

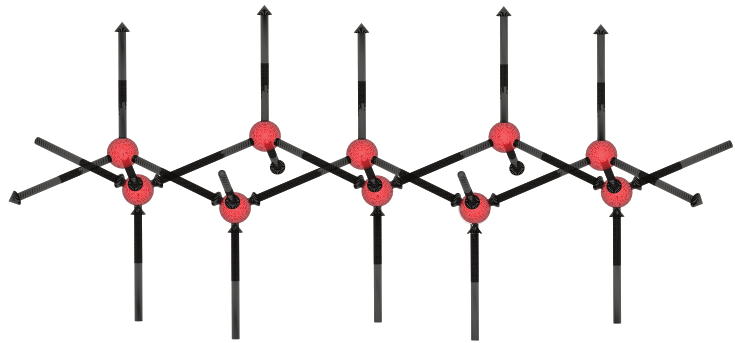

**Fig. 1**. The standard direction (+1) of the bonds in ice hexagonal monolayer. The oxygens are marked in red.

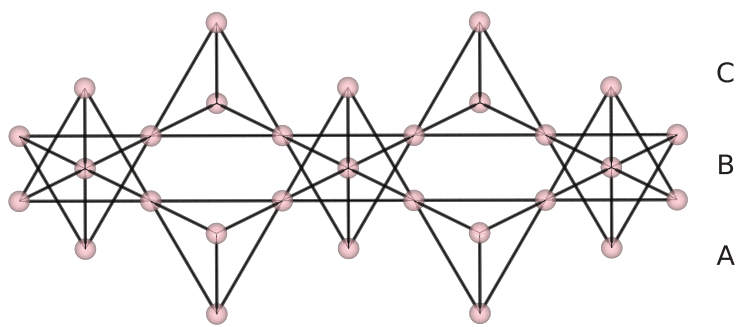

**Fig. 2**. The equivalent Ising model of ice hexagonal monolayer.

One can see from Fig. 2 that, the Ising lattice is in a form of "ABC", where "A" and "C" are two sparse triangular layers and "B" is the Kagomé layer. "B" layer corresponds to the hydrogen bond network and "A" and "C" layers correspond to the dangling bonds. Clearly, the configurations with two +1 and two -1 bonds around every oxygen are those obeying the ice rules. These configurations correspond to that in the Ising lattice with two +1 and two -1 spins around every tetrahedron. Then let us consider the spins on the Kagomé layer ("B" layer). When there are two +1 and two -1 spins around every tetrahedron, there must be two +1 and one -1 or two -1 and one +1 spins in every triangle of the Kagomé layer. This is exactly the ground states of the antiferromagnetic Ising model with nearest-neighbour interactions on the Kagomé lattice. That is, there is a one-to-one mapping of the configurations obeying the ice rules (in zero field) to the ground states of the antiferromagnetic Ising model on the Kagomé lattice. Notice that $N_{H_2O} = \frac{2}{3} N_K$, where $N_K$ is the number of spins on the Kagomé lattice. Then the residual entropy of ice hexagonal monolayer in zero field is simply obtained from that of the antiferromagnetic Ising model on the Kagomé lattice [36]

$$\begin{aligned} S/k_B &= \frac{3}{2} S_K/k_B \\ &= \frac{3}{2} \times \frac{1}{24\pi^2} \int_0^{2\pi} d\theta \int_0^{2\pi} d\phi \ \ln\{21 - 4[\cos\theta + \cos\phi + \cos(\theta+\phi)]\} \\ &= 0.752745 \ . \end{aligned} \qquad (1)$$

In the presence of an external electric field along the z-axis, the hydrogens in the dangling bonds are pined parallel to the field. Then, the configurations of the dangling bonds are constrained, i.e., all the spins in "A" and "C" layers are +1. In this case, obviously the residual entropy is reduced. The ice rules force the configurations to those with two -1 and one +1 spins in every triangle of "B" layer. The system under this condition is called the "Kagomé ice" [86]. The reduced but still extensive ground state degeneracy was exactly solved by Moessner *et al.* [87] by mapping the ground states to the configurations of the dimer model on the honeycomb lattice [67, 69]. They showed that the residual entropy is one half of that of the antiferromagnetic Ising model on the triangular lattice [34] $S = \frac{1}{2} S_{tri}$. Later, Matsuhira *et al.* measured this residual entropy in a pyrochlore spin ice material [86]. Udagawa *et al.* rediscovered the mapping to the dimer model on the honeycomb lattice, and rederived the solution using the Pfaffian method [88]. This residual entropy is also discussed by a few other studies, both in the spin ice system [11, 17, 20, 89, 90] and in real ice [91].

Here we will propose an alternative approach to this solution. Noticing the ground states in this case are those with two -1 and one +1 spins in every triangle of the Kagomé lattice, we consider the low temperature limit of the antiferromagnetic Ising model with nearest-neighbour and three-spin interactions on the Kagomé lattice. The interaction energy within each triangle sounded by three spins $(s_1, s_2, s_3)$ is

$$E(s_1, s_2, s_3) = J(s_1 s_2 + s_2 s_3 + s_1 s_3) + \Delta(s_1 s_2 s_3 - 1) , \qquad (2)$$

where $J > 0$ is the nearest-neighbour interaction and $\Delta$ is the three-spin interaction. The Hamiltonian of this Ising model can then be expressed as

$$H = \sum_{tri} E(s_1, s_2, s_3) , \qquad (3)$$

with the summation taken over all triangles. To exactly solve the partition function $Z$, we follow [92, 93] and map this Ising model into an eight-vertex model on the honeycomb lattice [94]. The vertex weights of this eight-vertex model are given by

$$\begin{aligned}
a &= \exp[-\beta E(1,1,1)] = \exp[-3\beta J], \\
b &= \exp[-\beta E(1,1,-1)] = \exp[-\beta(-J - 2\Delta)], \\
c &= \exp[-\beta E(1,-1,-1)] = \exp[\beta J], \\
d &= \exp[-\beta E(-1,-1,-1)] = \exp[-\beta(3J - 2\Delta)],
\end{aligned} \qquad (4)$$

where $\beta = 1/k_B T$. It is straightforward to verify that, in choosing $\Delta = -\infty$, the energy within a triangle is in the order: $E(1,-1,-1) < E(1,1,1) < E(1,1,-1) = E(-1,-1,-1) = +\infty$. The ground states are exactly the configurations with two -1 and one +1 spins in every triangle. By taking the low temperature limit, the residual entropy can then be obtained from the ground state degeneracy as

$$S/k_B = \lim_{N_{tri} \to \infty} \frac{1}{N_{tri}} \ln[g(E_0)] = \lim_{\beta \to \infty} \left\{ \lim_{N_{tri} \to \infty} \frac{1}{N_{tri}} (\ln Z + \beta E_0) \right\} \quad (5)$$

with the ground state energy $E_0 = N_{tri} \times E(1,-1,-1) = -N_{tri} J$. Here $N_{tri}$ is the number of triangles, and is also the number of vertices on the honeycomb lattice. It is easy to find $N_{H_2O} = N_{tri}$. Now we may make use of the equivalence with the eight-vertex model on the honeycomb lattice

$$Z = Z_{8v}(a,b,c,d) \quad (6)$$

in the case that $a = e^{-3\beta J}$, $c = e^{\beta J}$, $b = d = 0$. Fortunately, the eight-vertex model in this case is exactly solved in [94]. We recall Eq. (17) of [94], and substitute it into Eq. (5)

$$\begin{aligned}
&\lim_{N_{tri} \to \infty} \frac{1}{N_{tri}} (\ln Z + \beta E_0) \\
&= \lim_{N_{tri} \to \infty} \frac{1}{N_{tri}} \ln Z_{8v}(a,b,c,d) - \beta J \\
&= \frac{1}{16\pi^2} \int_0^{2\pi} d\theta \int_0^{2\pi} d\phi \ln\left\{ \left[ a^4 + 3c^4 + 2c^2(c^2 - a^2)(\cos\theta + \cos\phi + \cos(\theta + \phi)) \right] \times e^{-4\beta J} \right\}.
\end{aligned} \quad (7)$$

Inserting $a = e^{-3\beta J}$ and $c = e^{\beta J}$ into Eq. (7) gives the low temperature limit

$$\begin{aligned}
&\lim_{\beta \to \infty} \left[ \lim_{N_{tri} \to \infty} \frac{1}{N_{tri}} (\ln Z + \beta E_0) \right] \\
&= \frac{1}{16\pi^2} \int_0^{2\pi} d\theta \int_0^{2\pi} d\phi \ln\left[ 3 + 2(\cos\theta + \cos\phi + \cos(\theta + \phi)) \right] \\
&= \frac{1}{16\pi^2} \int_0^{2\pi} d\theta \int_0^{2\pi} d\phi \ln\left[ 1 + 4\cos\left(\frac{\theta+\phi}{2}\right)\cos\left(\frac{\theta-\phi}{2}\right) + 4\cos^2\left(\frac{\theta+\phi}{2}\right) \right] \\
&\stackrel{\sigma_1 = \frac{\theta+\phi}{2}, \sigma_2 = \frac{\theta-\phi}{2}}{=} \frac{1}{16\pi^2} \times 2 \times \int_\Omega d\sigma_1 d\sigma_2 \ln\left[ 1 + 4\cos\sigma_1 \cos\sigma_2 + 4\cos^2\sigma_1 \right].
\end{aligned} \quad (8)$$

Here the integral domain $\Omega$ is $\{(\sigma_1, \sigma_2): 0 \leq \sigma_1 + \sigma_2 \leq 2\pi \text{ and } 0 \leq \sigma_1 - \sigma_2 \leq 2\pi\}$. It is trivial to verify

$$\int_\Omega d\sigma_1 d\sigma_2 \ln\left[1 + 4\cos\sigma_1 \cos\sigma_2 + 4\cos^2 \sigma_1\right]$$
$$= \frac{1}{2} \int_0^{2\pi} d\sigma_1 \int_0^{2\pi} d\sigma_2 \ln\left[1 + 4\cos\sigma_1 \cos\sigma_2 + 4\cos^2 \sigma_1\right]. \quad (9)$$

Then Eq. (8) becomes

$$\lim_{\beta \to \infty}\left[\lim_{N_{tri} \to \infty} \frac{1}{N_{tri}} (\ln Z + \beta E_0)\right] = \frac{1}{16\pi^2} \int_0^{2\pi} d\sigma_1 \int_0^{2\pi} d\sigma_2 \ln\left[1 + 4\cos\sigma_1 \cos\sigma_2 + 4\cos^2 \sigma_1\right]. \quad (10)$$

Now we examine the residual entropy of the antiferromagnetic Ising model on the triangular lattice. In Page 364 of [34], this entropy had been shown in the form of

$$S_{tri}/k_B = \frac{1}{8\pi^2} \int_0^{2\pi} d\omega \int_0^{2\pi} d\omega' \ln\left[1 - 4\cos\omega \cos\omega' + 4\cos^2 \omega'\right] \quad (11)$$

before the final expression. One can easily see that the integral in Eq. (10) is equal to that in Eq. (11). Hence, the exact solution of the residual entropy in this case is exactly one half of that of the antiferromagnetic Ising model on the triangular lattice

$$S/k_B = \frac{1}{2} S_{tri}/k_B = 0.161533 \,. \quad (12)$$

Then we have rederived the result of [87] and [88], by a different approach of taking the low temperature limit of an Ising model on the Kagomé lattice.

We remark that, if we consider the Ising model on the Kagomé lattice with nearest-neighbour interactions $J > 0$ and an external magnetic field $H_{ex} = 4J$, the ground states will be the configurations with two -1 and one +1 or three -1 spins in every triangle. Even in the ground states, the triangles with three -1 spins will disobey the ice rules. This model is equivalent to the monomer-dimer model on the honeycomb lattice [11, 20, 79], where the triangles with two -1 and one +1 spins can be seen as the sites covered by a dimer, and those with three -1 spins can be seen as monomers. Solving the monomer-dimer model is much more difficult than the dimer covering problem. Nagle obtained the Bethe approximation for the monomer-dimer model on the honeycomb lattice [75], which was rederived by Isakov *et al*. [11]. Other theoretical approximation [90] and numerical results by Wang-Landau Monte Carlo simulation [20] were also studied. The exact solution of the residual entropy in this case has yet to be obtained.

## 3. Ice Hexagonal Monolayer under Periodic Boundary Conditions

We consider in this case, the ice hexagonal monolayer in the lattice of ice Ic. Three-dimensional model of ice Ic is equivalent to the pyrochlore antiferromagnetic Ising model [37]. The residual entropy of this model has been evaluated from various approaches [11, 16-18, 20, 23, 26, 90, 95, 96]. Here we study the ice hexagonal monolayer as a two-dimensional structure from this model.

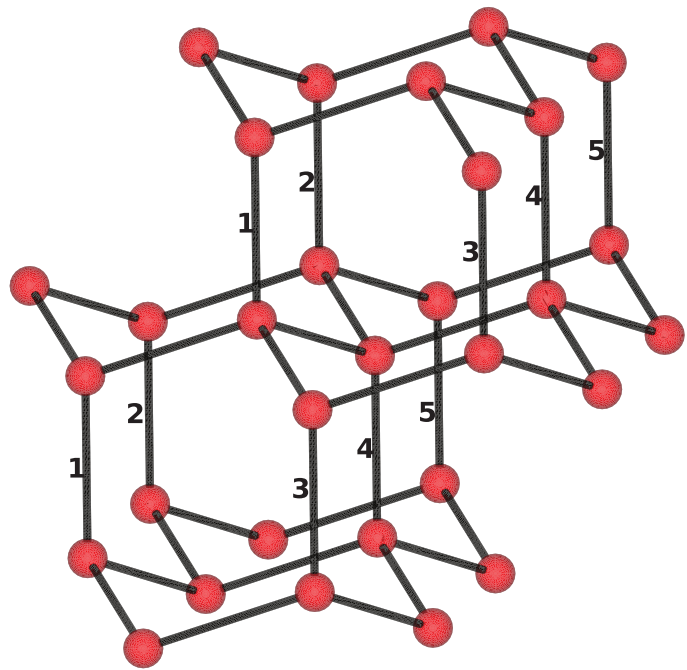

**Fig. 3**. The diagrammatic representation of the ice hexagonal monolayer in the lattice of ice Ic. The oxygen atoms are marked in red. Five "pairs" of dangling bonds under the periodic boundary conditions are marked.

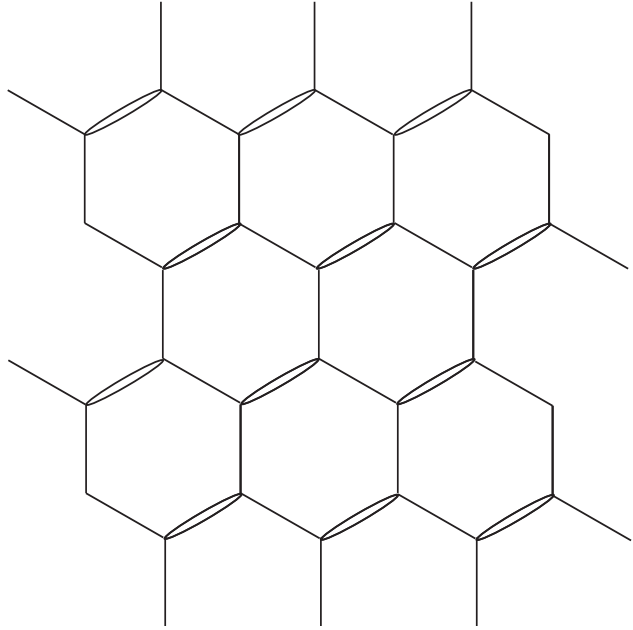

**Fig. 4**. The hydrogen bond network of the ice hexagonal monolayer under the periodic boundary conditions in ice Ic.

For the ice hexagonal monolayer in ice Ic, the periodic boundary conditions are taken on the dangling bonds of the layer. In Fig. 3, we show the example of five "pairs" of dangling bonds under the periodic boundary conditions. It is clear to verify that, each "pair" of dangling bonds are located on one pair of nearest-neighbour oxygens, and convert into one hydrogen bond under the periodic boundary conditions. Actually, there are double bonds connecting these pairs of nearest-neighbour oxygens. We show the two-dimensional hydrogen bond network in this case in Fig. 4.

The residual entropy of this model, can be exactly solved by employing the mapping into a six-vertex model on the square lattice. Consider the direction of each hydrogen bond in the network as an arrow. According to the ice rules, four arrows around each oxygen should be two-in-two-out respective to this oxygen. In Fig. 4 one can easily see that, each pair of nearest-neighbour oxygens connected by double bonds can be seen as a single site on the square lattice.

It is trivial to find that, the double bonds already contribute two-in-two-out, of the total four-in-four-out to these two oxygens, then the four arrows around this site should be two-in-two-out respective to this site. That is, the arrow configurations on the square lattice should be two-in-two-out respective to every site. We then obtain a six-vertex model on the square lattice equivalent to the hydrogen bond network, as shown in Fig. 5.

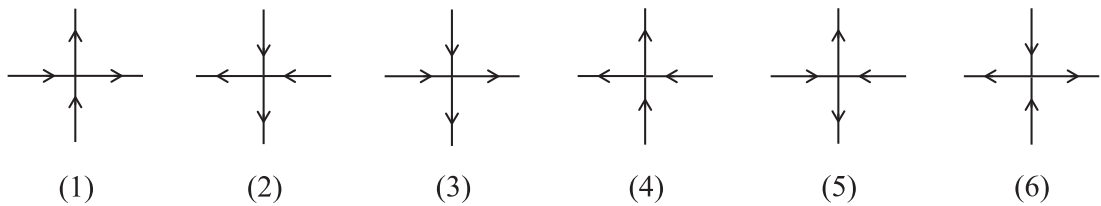

(1)　　　　　(2)　　　　　(3)　　　　　(4)　　　　　(5)　　　　　(6)

**Fig. 5**. The arrow configurations of the six-vertex model.

As shown in Fig. 5, the configuration of vertex (1) allows one hydrogen configuration in the double bonds of this site, and so does vertex (2). Hence the weights of (1) and (2) should be 1. Similarly, the weights of (3), (4), (5) and (6) should be 2, as there are two possibilities for the configuration of the double bonds. We list all the vertex weights

$$\omega_1 = \omega_2 = 1, \ \omega_3 = \omega_4 = \omega_5 = \omega_6 = 2 \ . \tag{13}$$

The residual entropy is then determined directly by the partition function of this model $S/k_B = \frac{1}{2} \times \lim_{N_{6v} \to \infty} \frac{1}{N_{6v}} \ln Z$, where $N_{6v}$ is the number of vertices on the square lattice and $N_{H_2O} = 2N_{6v}$. Remark that the six-vertex model in the case $\omega_1 = \cdots = \omega_6 = 1$ is the square ice model. To solve our six-vertex model, we should first recall the exactly solved cases in the previous studies. In [43-45], the solution of a six-vertex model with the vertex weights and the associated energies

$$\varepsilon_1 = \varepsilon_2 = -\frac{1}{2}\delta, \ \tilde{\omega}_1 = \tilde{\omega}_2 = e^{\frac{1}{2}\beta\delta},$$

$$\varepsilon_3 = \varepsilon_4 = \frac{1}{2}\delta, \ \tilde{\omega}_3 = \tilde{\omega}_4 = e^{-\frac{1}{2}\beta\delta}, \quad (14)$$

$$\varepsilon_5 = \varepsilon_6 = -\varepsilon, \ \tilde{\omega}_5 = \tilde{\omega}_6 = e^{\beta\varepsilon}$$

and in the case $\delta > 0$ was examined. Notice that the solution is invariant under a C$_2$ symmetry operation along the vertical axis, i.e., under the transformation $\tilde{\omega}_1 \leftrightarrow \tilde{\omega}_4$, $\tilde{\omega}_2 \leftrightarrow \tilde{\omega}_3$ (see Fig. 5).

We may set the vertex energies as $\beta\delta = \ln 2$ and $\varepsilon = \frac{1}{2}\delta$, and redefine the vertex weights as

$$\tilde{\omega}_1 = \tilde{\omega}_2 = e^{-\frac{1}{2}\beta\delta} = \frac{1}{\sqrt{2}}, \ \tilde{\omega}_3 = \tilde{\omega}_4 = \tilde{\omega}_5 = \tilde{\omega}_6 = e^{\frac{1}{2}\beta\delta} = \sqrt{2}. \quad (15)$$

Then the relation of the partition function of our model with that of the model defined in Eq. (15) is straightforward

$$Z = \sqrt{2}^{N_{6v}} \times \tilde{Z}. \quad (16)$$

To evaluate the solution of $\tilde{Z}$, we follow the work of [44, 45, 97] and determine the quantities

$$\eta = e^{\beta\delta} = 2, \ \xi = e^{2\beta\varepsilon} = 2,$$

$$\Delta = \frac{1}{2}(\eta + \eta^{-1} - \xi) = \frac{1}{4}. \quad (17)$$

Define $\mu$ and $\Phi_0$ by

$$\cos\mu = -\Delta,$$

$$e^{i\Phi_0} = \frac{1 + \eta e^{i\mu}}{\eta + e^{i\mu}}. \quad (18)$$

It is simple to give

$$\mu = \arccos\left(-\frac{1}{4}\right), \ \Phi_0 = \arccos\left(\frac{11}{16}\right). \quad (19)$$

Now we recall Eq. (15) of [45], which is the exact solution of $\tilde{Z}$, in the case that the vertical polarization and horizontal polarization are 0. The vertical and horizontal polarizations are associated with the external fields in the vertical and horizontal directions, respectively. In our case, they are certainly 0. Then we obtain

$$\lim_{N_{6v} \to \infty} \frac{1}{N_{6v}} \ln Z = \frac{1}{2} \ln 2 + \lim_{N_{6v} \to \infty} \frac{1}{N_{6v}} \ln \tilde{Z}$$

$$= \frac{1}{2} \ln 2 + \frac{1}{2} \beta \delta + \frac{1}{8\mu} \int_{-\infty}^{\infty} \frac{d\alpha}{\cosh(\pi\alpha/2\mu)} \ln \left[ \frac{\cosh \alpha - \cos(2\mu - \Phi_0)}{\cosh \alpha - \cos \Phi_0} \right] \quad (20)$$

$$= \frac{1}{2} \ln 2 + \frac{1}{2} \beta \delta + \frac{1}{4} \int_{-\infty}^{\infty} \frac{d\alpha}{\cosh(\pi\alpha)} \ln \left[ \frac{\cosh(2\mu\alpha) - \cos(2\mu - \Phi_0)}{\cosh(2\mu\alpha) - \cos \Phi_0} \right]$$

$$= 0.946954 .$$

The exact residual entropy is

$$S/k_B = \frac{1}{2} \times \lim_{N_{6v} \to \infty} \frac{1}{N_{6v}} \ln Z = 0.473477 . \quad (21)$$

Interestingly, in [24] Kirov constructed a digonal hexagonal ice model, of which the hydrogen bond network is very similar to that of our model (see Fig. 5(b) of [24]). Kirov developed a numerical transfer matrix method to completely enumerate the hydrogen bond configurations in finite fragments of various models [24, 25]. For the digonal hexagonal ice model, his result of the largest fragment is 0.474497, which is very close to the solution of our model. It is not clear whether the exact residual entropies in the large lattice limit of these two models are consistent, though.

## 4. Discussions and Conclusions

In this work we study the exact residual entropy of ice hexagonal monolayer, a two-dimensional structure from real ice. We have examined two soluble cases. The case in the presence of an external electric field along the z-axis, also called as the "Kagomé ice", has been exactly solved in the previous studies [87-89]. The model in this case is equivalent to the dimer model on the honeycomb lattice, and the residual entropy is equal to the solution of the dimer covering problem, which is one half of the residual entropy of the antiferromagnetic Ising model on the triangular lattice. We give an alternative approach to this solution by mapping the model into an Ising model with nearest-neighbour and three-spin interactions on the Kagomé lattice. The ground states of this Ising model are exactly the configurations in the "Kagomé ice", hence we take the low temperature limit of this Ising model and obtain the residual entropy. We show that this result is exactly one half of that of the antiferromagnetic Ising model on the triangular lattice, therefore finish the rederivation. The advantage of making use of the equivalence of the ice-type model with the Ising spin model is demonstrated in our method.

The second case we consider is the ice hexagonal monolayer under the periodic boundary conditions in ice Ic. In this case each "pair" of dangling bonds located on one pair of nearest-neighbour oxygens convert into a hydrogen bond, then this pair of nearest-neighbour oxygens are connected by double bonds. The hydrogen bond network can then be mapped into a six-vertex model on the square lattice, with each pair of nearest-neighbour oxygens connected by double bonds seen as a single site. We transform this six-vertex model to an exactly solved case [43-45], and obtain the solution of residual entropy. It is worth comparing this solution with the result of three-dimensional ice Ic. We recall the estimates of the residual entropy of ice Ic in [23] and [26], which are the most accurate series expansion result and theoretical result, respectively. Either the series expansion result 0.411014, or the theoretical result $\ln(1.507456) = 0.410423$, has a significant difference with the exact result of the hexagonal monolayer 0.473477 [see Eq. (21)]. This fact confirms that, the correlation between layers in three-dimensional ice restricts the hydrogen bond configurations. Anyway, our solution for the hexagonal monolayer under the periodic boundary conditions provides the result of a very simplified version of real ice. Effects of the correlation between layers in three-dimensional ice deserve further study.

In conclusion, for the ice hexagonal monolayer, we rederive the result in a solved case and give the solution in a new soluble case. In this article we provide more examples of the research of residual entropy problem in various ice systems. Our work widens the set of exactly soluble two-dimensional models in statistical physics.

## Statements and Declarations

**Funding**

This work was supported by Guangdong Basic and Applied Basic Research Foundation (Grant No. 2021A1515010328), Key-Area Research and Development Program of Guangdong Province (Grant No. 2020B010183001), and National Natural Science Foundation of China (Grant No. 12074126).

**Competing interests**

The authors have no competing interests to declare that are relevant to the content of this paper.

## Data Availability Statements

Data sharing is not applicable to this paper as no datasets were generated or analysed during the current study.